\documentclass[11pt]{article}
\pdfoutput=1
\usepackage{jcappub}
\usepackage{graphicx}

\usepackage[dvipsnames]{xcolor}
\usepackage{xspace}

\usepackage[normalem]{ulem}

\usepackage{subcaption}

\title{Light bending by the cosmological constant}
\author[a]{Lingyi Hu}
\author[a]{Alan Heavens}
\author[b]{David Bacon}

\affiliation[a]{Imperial Centre for Inference and Cosmology (ICIC), Imperial College, London, SW7 2AZ, U.K.}
\affiliation[b]{Institute of Cosmology and Gravitation, University of Portsmouth, 1-8 Burnaby Road, Portsmouth PO1 3FX, U.K.}

\emailAdd{lingyihuu@gmail.com}
\emailAdd{a.heavens@imperial.ac.uk}
\emailAdd{david.bacon@port.ac.uk}

\keywords{}

\abstract{We revisit the question of whether the cosmological constant $\Lambda$ affects the cosmological gravitational bending of light, by numerical integration of the geodesic equations for a Swiss cheese model consisting of a point mass and a compensated vacuole, in a Friedmann-Robertson-Walker background. We find that there is virtually no dependence of the light bending on the cosmological constant that is not already accounted for in the angular diameter distances of the standard lensing equations, plus small modifications that arise because the bending is restricted to a finite region covered by the hole.  The residual $\Lambda$ dependence for a $10^{13}\,M_{\odot}$ lens is at the level of 1 part in $10^7$, and even this might be accounted for by small changes in the hole size evolution as the photon crosses. We therefore conclude that there is no need for modification of the standard cosmological lensing equations in the presence of a cosmological constant.}

\begin{document}

\maketitle
\flushbottom

\section{Introduction}

An active debate that has been the subject of many papers in the last decade is to what extent, if any, the cosmological constant $\Lambda$ directly affects gravitational lensing. Standard lensing formalism already takes into account an implicit dependence through the dependence on $\Lambda$ of angular diameter distances, and the question is if any modification to the current lensing formalism is needed in a universe with a positive $\Lambda$. The conventional view, first put forth by Islam \citep{Islam1993}, is that no such modification is necessary, due to the straightforward observation that $\Lambda$ does not appear in the null geodesic equations of a photon orbit in Kottler spacetime. This view was first challenged by Rindler and Ishak \citep{Rindler2007} who argue that while $\Lambda$ drops out of the equations of motion, it still affects light bending through the metric of spacetime itself, since the photon is moving in a $\Lambda$-dependent geometry. Since then, there have been many papers that investigate the influence of the cosmological constant on gravitational lensing \cite{Gibbons2008, Finelli2007, Sereno2008, Schucker2008, Bhadra2010, Piattella2016, Aghili2017, Khriplovich2008, Park2008, Simpson2010, Arakida2012, Butcher2016},  of which several papers have supported and built on Rindler and Ishak's results \citep{Schucker2008, Bhadra2010, Piattella2016, Aghili2017}. That $\Lambda$ influences the bend angle through its effect on angular diameter distances is not in doubt (e.g., \cite{Sereno2008, Simpson2010}); the question is whether there are other effects, and a number of papers have argued against this \citep{Khriplovich2008, Park2008, Simpson2010, Arakida2012, Butcher2016}.

A major difference between the Kottler and Schwarzschild spacetimes is that in the former, spacetime is not asymptotically flat, so some of the standard lensing arguments regarding angles and distances do not apply. Previous work used a variety of approaches to take this into account, for example, \cite{Arakida2012, Butcher2016} considered a Kottler spacetime with modifications in the angle calculations, \cite{Park2008, Piattella2016, Aghili2017} used a McVittie metric, while \cite{Kantowski2010, Schucker2008, Schucker2009} worked with a Swiss-cheese model that stitches together the Kottler and Friedmann-Robertson-Walker (FRW) spacetimes. 

In this work, we aim to isolate the effect of $\Lambda$ through numerical integration of light propagation in a Swiss-cheese model, where the lens is enclosed in a spherical void (a `Kottler hole') with a size chosen to compensate exactly for the mass concentrated at the centre. The hole is embedded into a uniform background FRW spacetime. This is clearly not a realistic structure model, but as a non-trivial test case, it has the great advantage that the evolution of the matter distribution is known exactly. In more general situations, the matter evolution needs to be computed simultaneously with the photon trajectory. However, this case is sufficiently rich to produce different predictions for the bending from the different approaches in the literature, such that general conclusions are able to be drawn. As the observer and source are both in the FRW region, we can use standard angular diameter distance formulae to calculate observed angles and distances as we would in conventional lensing, and bending only happens in a finite region inside the hole. The expected corrections to the standard gravitational lensing formalism in the Swiss-cheese solution arising from the finite size of the hole were calculated in \citep{Kantowski2010}. Schucker \cite{Schucker2008, Schucker2009} has done some similar work in numerical simulations of the Swiss-cheese, and concluded that he agrees with Rindler and Ishak. We reproduce his results and expand upon it, but come to the opposite conclusion, and explain some of the apparent discrepancies. 

Some of the controversy may be due to the many factors involved when changing $\Lambda$, since a change in $\Lambda$ affects the whole cosmology, for example:

\begin{itemize}
  \item In flat space, increasing $\Lambda$ reduces the matter density, causing the size of the Kottler hole to increase, assuming the lens mass is kept constant.
  \item When the redshift to the source is fixed, increasing $\Lambda$ increases the angular diameter distance to the source.
  \item The rate of expansion of the Kottler hole in static coordinates changes with $\Lambda$, as a result of the matching conditions between the two metrics at the boundary.
\end{itemize}

What we seek here are effects of $\Lambda$ that are not taken account of in the standard lensing formulae, since the main practical goal here is to know if inferences from gravitational lensing are likely to be inaccurate because of the presence of $\Lambda$. There are some modifications, such as those above, and the fact that outside the hole the geodesics are not perturbed by the lensing mass.

In this paper we control for these effects, and account for these indirect contributions in order to isolate any residual effects of $\Lambda$, so as to determine whether a correction term involving $\Lambda$ is necessary. We find that the residual effects of $\Lambda$ are extremely small, at the level of 1 part in $10^7$ for a $10^{13}\,M_\odot$ lens ($10^{-5}$ for $M=10^{15}\,M_\odot$, and even this may possibly be accounted for by small changes in the evolution of the size of the hole during the passage of the photon. 

\section{The Swiss cheese model with a cosmological constant}
\label{sec:swiss-cheese}

We use a Swiss-cheese model to model light bending in an expanding universe. Such a model is constructed by evacuating the matter from a comoving sphere in the homogeneous background and replacing it with an inhomogeneous mass distribution. In our case this will be vacuum everywhere except for a point mass at the centre, with a non-zero cosmological constant. The Friedmann-Robertson-Walker (FRW) metric describes the geometry outside the hole and the Kottler metric describes the geometry inside the hole. The advantage of using this model is that we already know the solution to the background evolution of the matter, and do not need to solve for it in parallel with the light path integrations.  For simplicity, we choose the observer, lens, and source to be collinear.  This simple case is sufficient to seek residual $\Lambda$ effects.

\begin{figure}
  \centering
  \includegraphics[height=0.4\linewidth]{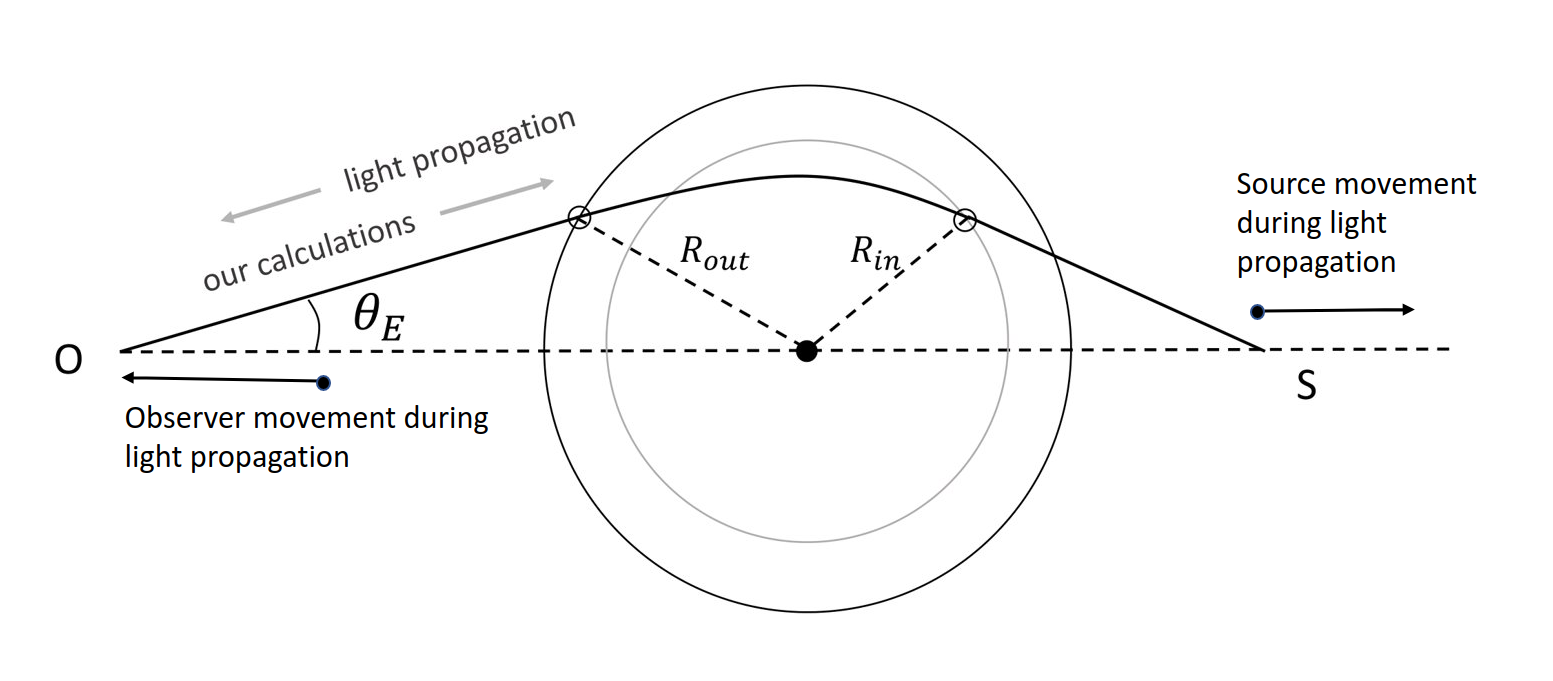}
  \caption{A diagram of how a light ray propagates through the Swiss-cheese. Our calculations are done in the opposite direction of propagation. Due to the expansion of the universe, while the hole size stays the same in comoving coordinates, the physical size of the hole will be larger when the light exits the hole compared to when it entered. During the light propagation, both the source and observer will have moved apart in proper coordinates, but they remain in the same position in comoving coordinates. Light rays that miss the hole follow the homogeneous universe geodesics, since the point mass is compensated by the evacuation of the hole.}
  \label{fig:swiss-cheese-ray}
\end{figure}

We propagate a light ray backwards from observer to source by solving the null geodesic equations in each region numerically. We convert from FRW to Kottler metric and vice versa at the boundary using the junction conditions. The observer is in the FRW region, the light ray starts with an Einstein angle $\theta_{E}$, travels through the FRW region, encounters the Kottler hole which deflects its trajectory, and returns to the FRW region again (see \autoref{fig:swiss-cheese-ray}). Since the observer and source are at rest in comoving coordinates, no correction for aberration is needed.

\subsection{FRW region}
Outside the hole, the geometry is described by the FRW metric (we take $c=G=1$ throughout)

\begin{equation}
  ds^2 = -dt^2 + a(t)^2 \left ( \frac{dr^2}{1-kr^2} + r^2 d \Omega^2 \right )
  \label{eq:frw-metric}
\end{equation}
where $d \Omega^2 = d\theta^2 + \sin^2\theta d\phi^2$ is the metric on a 2-sphere, $a(t)$ is the scale factor, and $k$ represents the curvature of space. The scale factor $a(t)$ satisfies the Friedmann equation

\begin{equation}
  H^2(a) \equiv \left ( \frac{1}{a}\frac{da}{dt} \right )^2 = \frac{8\pi \rho}{3} + \frac{\Lambda}{3} - \frac{k}{a^2}
  \label{eq:friedmann-equation}
\end{equation}
where $\rho$ is the energy density of a pressureless fluid and $H(a)$ is the Hubble parameter. It is common to introduce the cosmological parameters, where a subscript 0 refers to quantities evaluated today: 

\begin{equation}
  \Omega_{\rm m} = \frac{8\pi \rho_0}{3H_0^2}, \,\, \Omega_{\Lambda} = \frac{\Lambda}{3H_0^2}, \,\, \Omega_k = - \frac{k}{a_0^2 H_0^2}
  \label{eq:cosmo-params}
\end{equation}
and rewrite the Friedmann equation as

\begin{equation}
  H^2 = H_0^2 \left [ \Omega_{\rm m} \left ( \frac{a_0}{a}\right )^3 + \Omega_k \left ( \frac{a_0}{a}\right )^2 + \Omega_{\Lambda} \right ].
  \label{eq:friedmann-eqn-version2}
\end{equation}
At the present day, ignoring radiation, the density parameters obey the relation

\begin{equation}
  \Omega_{\rm m} + \Omega_{\Lambda} + \Omega_k = 1.
  \label{eq:density-parameters-1}   
\end{equation} 
Applying the geodesic equations that govern the trajectory of the light ray in the FRW region, we find
\begin{subequations}
  \begin{align}
    \dot{t} &= -\sqrt{\frac{a^2\dot{r}^2}{1-kr^2} + a^2r^2 \dot{\phi}}\\
    \ddot{r}  &= (1-kr^2)r\dot{\phi}^2 - \frac{kr\dot{r}^2}{1-kr^2} - \frac{2a_{,t}}{a}\dot{r}\dot{t}\\
    \dot{\phi} &= \frac{L}{a^2 r^2}
  \end{align}
  \label{eq:frw-null-geodesics}%
\end{subequations}
where $L = a^2 r^2 \dot{\phi}$ is a conserved quantity, an overdot denotes derivative with respect to an affine parameter, and comma denotes a partial derivative. When combined with the Friedmann equation (\autoref{eq:friedmann-eqn-version2}), they fully determine the light's path. The negative sign on $\dot{t}$ is due to the fact that we are propagating the light backwards in time. We can then solve these differential equations numerically and stop the integration once the light ray reaches the boundary of the hole, which is defined by $r_{\rm h} = \text{constant}$. At the boundary, we apply junction conditions (discussed below in \autoref{section:boundary}) to convert the FRW coordinates into Kottler coordinates. We continue the integration inside the Kottler vacuole until the boundary is reached, and we apply the junction conditions again to convert Kottler coordinates back into FRW coordinates. We continue the integration in the FRW region and stop once it reaches the $x$-axis.

\begin{figure}
  \centering
  \includegraphics[height=0.2\linewidth]{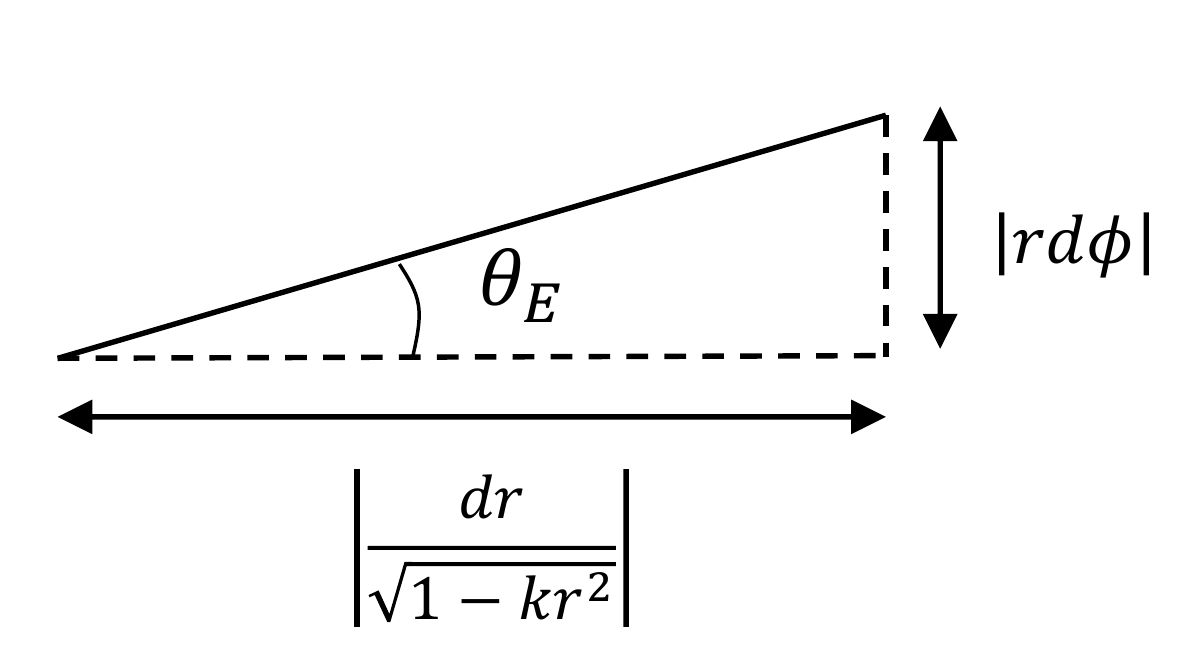}
  \caption{The physical angle is found by restricting the FRW metric to $t = \text{constant}$. With $\theta = \pi/2$ and $a_0 = 1$, the metric becomes $ds^2 = dr^2 / (1-kr^2) + r^2 d\phi^2$. The light path forms the hypotenuse of the infinitesimal triangle in the diagram. }
  \label{fig:angle-to-tangent-vectors}
\end{figure}

Since we integrate the equations backwards, we begin by fixing the observed angle of the image with respect to the lens, which will be the Einstein angle $\theta_{\rm E}$, since we have a collinear arrangement. We fix the position of the lens by specifying its angular diameter distance $D_{\ell}$. Our goal is to deduce the position of the source which gives rise to the image, and with everything then determined, compare with predicted formulae.   From $D_{\ell}$, we obtain the comoving distance $r_{\ell}$ of the lens and the initial tangent vectors of the light ray to start off the integration. We place the lens at the origin and take the observer to be at an azimuthal angle of $\phi=\pi$, so the initial tangent vectors are related to $\theta_{\rm E}$ by (see \autoref{fig:angle-to-tangent-vectors})

\begin{equation}
  \theta_{\rm E} = \tan^{-1}\left (\frac{\sqrt{1-kr^2}\,r\dot{\phi}}{\dot{r}}\right ).
  \label{eq:angle-to-tangent-vectors}
\end{equation}

Assuming $a_0 = 1$, the angular diameter distance relates to the redshift by \citep{Hogg1999}

\begin{equation}
  D_{\ell} = D_{\rm A}(z_{\ell}) = \frac{1}{1+z_{\ell}} S_k\left[ \frac{1}{H_0}\int_0^{z_{\ell}} \frac{dz}{\sqrt{\Omega_{\rm m}(1+z)^3 + \Omega_k (1+z)^2 + \Omega_{\Lambda}}}  \right]
  \label{eq:angular-diameter-distance-redshit-z}
\end{equation}
where we define the function $S_k(x)$

\begin{equation}
  S_k(x) \equiv
  \begin{cases}
    \mathopen| k \mathclose|^{-1/2} \sin(\sqrt{\mathopen| k \mathclose|}x) & k > 0\\
    x & k = 0\\
    \mathopen| k \mathclose|^{-1/2} \sinh(\sqrt{\mathopen| k \mathclose|}x) & k < 0
  \end{cases}
  \label{eq:sk}
\end{equation}
and its inverse function

\begin{equation}
  S_k^{-1}(x) \equiv 
  \begin{cases}
    \mathopen| k \mathclose|^{-1/2} \sin^{-1}(\sqrt{\mathopen| k \mathclose|}x) & k > 0\\
    x & k = 0 \\
    \mathopen| k \mathclose|^{-1/2} \sinh^{-1}(\sqrt{\mathopen| k \mathclose|}x) & k < 0.
  \end{cases}
  \label{eq:sk-inverse}
\end{equation}
The radial coordinate is given by 
\begin{equation}
  r_{\ell} = (1+z_{\ell})D_{\ell}.
\end{equation}
As the lens is placed at the centre, the observer is placed at a comoving distance of $r = r_{\ell}$. 

\subsection{Kottler region}

Inside the hole, the metric is a static Kottler metric \citep{Kottler1918}, with radial coordinate $R$ and time coordinate $T$, and is given by

\begin{equation}
  ds^2 = -f(R)dT^2 + \frac{dR^2}{f(R)} + R^2 d \Omega^2
  \label{eq:kottler-metric}
\end{equation}
with
\begin{equation}
  f(R) = 1-\frac{2M}{R} - \frac{\Lambda R^2}{3},
  \label{eq:kottler-metric-f}
\end{equation}
where $M$ is the mass of the central object. The Kottler hole has to be mass compensating, so $M$ has to equal the mass excised from the FRW background:

\begin{equation}
  M = \frac{4\pi}{3} \rho a^3 r_{\rm h}^3. 
  \label{eq:junction-conditions-mass-volume}
\end{equation}
where $r_{\rm h}$ is the comoving radius of the hole in FRW coordinates.

The Kottler metric has 2 conserved quantities, $E = f(R) \dot{T}$ and $L = R^2 \dot{\phi}$, and we can write the null geodesics as 

\begin{subequations}
  \begin{align}
    \dot{T} &= \frac{E}{f(R)}\\
    \ddot{R}  &= \frac{L^2 (R-3M)}{R^4}\\
    \dot{\phi} &= \frac{L}{R^2}.
  \end{align}
  \label{eq:kottler-null-geodesics}%
\end{subequations}
At the same time that the light ray is moving through the Kottler hole, the size of the hole $r_{\rm h}$ is also changing in the static coordinates, with an expansion rate that can be derived from the matching conditions between the two metrics.

\subsection{Matching conditions at the boundary}
\label{section:boundary}

We match the FRW and Kottler metric on a surface of a comoving 2-sphere which is defined by $r = r_{\rm h} = \text{constant}$ in FRW coordinates and $R = r_{\rm h}(T)$ in Kottler coordinates. The two geometries can be matched across the boundary via a hypersurface $\Sigma$ to form a well defined spacetime if they satisfy the Darmois-Israel junction conditions \citep{Israel1966}, which dictate that they must induce the same metric and extrinsic curvature at the boundary.  Matching needs to be done with some care, since we are not only concerned with the smoothness of the physical conditions, but also the smoothness with which the coordinates describe the space-time manifold. In this section we follow the development of \cite{Fleury2013} and \cite{Dupuy2015}, where a similar derivation was done.

The induced metric is the quantity 

\begin{equation}
  h_{ab} = g_{\alpha \beta} j^{\alpha}_{a} j^{\beta}_{b}
  \label{eq:induced-metric-defn}
\end{equation}
where $j^{\alpha}_{a}$ is defined as

\begin{equation}
  j^{\alpha}_{a} = \frac{\partial \bar{X}^{\alpha}}{\partial \sigma^a}.
  \label{eq:j-defn}
\end{equation}
The hypersurface $\Sigma$ is the world sheet of a comoving 2-sphere at the junction between the FRW and Kottler metric. Here we follow the treatment in \cite{Dupuy2015}, introducing $X^{\alpha}$ to represent coordinates of the original metric, $\sigma^a$ to be natural intrinsic coordinates for $\Sigma$, and $\bar{X}^{\alpha}(\sigma^a)$ is the parametric equation of the hypersurface. More concretely, using the coordinates defined previously in \autoref{eq:frw-metric}, these quantities are, for the FRW, 

\begin{subequations}
  \begin{align}
    X^{\alpha} &= \{ t, r, \theta, \phi \} \\
    \sigma^a &= \{ t, \theta, \phi \} \\
    \bar{X}^{\alpha}(\sigma^a) &= \{ t, r_{\rm h}, \theta, \phi\}.
  \end{align}
\end{subequations}
Similarly, in the Kottler region, we have 

\begin{subequations}
  \begin{align}
    X^{\alpha} &= \{ T, R, \theta, \phi \} \\
    \sigma^a &= \{ T, \theta, \phi \} \\
    \bar{X}^{\alpha}(\sigma^a) &= \{ T, r_{\rm h}(T), \theta, \phi\}.
  \end{align}
\end{subequations}
Using these definitions, the 3-metric induced by the FRW geometry on $\Sigma$ is

\begin{equation}
  ds^2_{\Sigma} = -dt^2 + a^2(t)r^2 d \Omega^2,
  \label{eq:frw-induced-metric}
\end{equation}
while the induced metric on the Kottler metric is
\begin{equation}
  ds_{\Sigma}^2 = -\kappa^2(T)dT^2 + r_{\rm h}^2(T) d \Omega^2,
  \label{eq:kottler-induced-metric}
\end{equation}
where
\begin{equation}
  \kappa \equiv \sqrt{\frac{f^2[r_{\rm h}(T)] - {r_{\rm h}^{\prime}}^2}{f[r_{\rm h}(T)]}}.
  \label{eq:kottler-kappa}
\end{equation}
where $r_{\rm h}^{\prime} = \frac{dr_{\rm h}(T)}{dT}$. Equating the components of \autoref{eq:frw-induced-metric} and \autoref{eq:kottler-induced-metric}, we obtain the following:

\begin{equation}
  r_{\rm h}(T) = a(t)r_{\rm h}
  \label{eq:r-to-ar},
\end{equation}

\begin{equation}
  \frac{dt}{dT} = \kappa(T).
  \label{eq:dt-dT}
\end{equation}

The second condition equates extrinsic curvature of the two geometries. By definition, the extrinsic curvature $K_{ab}$ of a hypersurface is given by
\begin{equation}
  K_{ab} = n_{\alpha;\beta} j^{\alpha}_{a} j^{\beta}_{a}
  \label{eq:extrinsic-curvature-defn}
\end{equation}
where $n_{\mu}$ is the unit vector normal to $\Sigma$, $j$ is as defined previously in \autoref{eq:j-defn}, and the semicolon notation `;' denotes a covariant derivative, $n_{\alpha;\beta} = \nabla_{\beta}\, n_{\alpha}$. For any vector $V^{\nu}$, the covariant derivative is defined as

\begin{equation}
  \nabla_{\mu}V^{\nu} = \partial_{\mu}V^{\nu} + \Gamma^{\nu}_{\mu \rho} V^{\rho}.
  \label{eq:covariant-derivative-defn}
\end{equation}

For a hypersurface defined by a function $q = 0$, the unit vector normal to it is

\begin{equation}
  n_{\mu} = \frac{q_{,\mu}}{\sqrt{g^{\alpha \beta} q_{,\alpha} q_{,\beta}}}.
  \label{eq:unit-normal-vector}
\end{equation}

In our case $q = r-r_{\rm h}$ in FRW coordinates and $q = R - r_{\rm h}(T)$ in Kottler coordinates. Using \autoref{eq:unit-normal-vector}, the unit vector in the FRW region is $n_{\mu}^{\text{(FRW)}} = a(t) \delta^r_{\mu}/\sqrt{1-kr^2}$. Applying \autoref{eq:extrinsic-curvature-defn}, the extrinsic curvature induced by the FRW geometry is

\begin{equation}
  K_{ab} dx^a dx^b = a(t)r \sqrt{1-kr^2} d \Omega^2
  \label{eq:extrinsic-curvature-frw}
\end{equation}
while the extrinsic curvature induced by the Kottler geometry is
\begin{equation}
  K_{ab} dx^a dx^b = \frac{1}{\kappa} \left [ r_{\rm h}^{\prime\prime} + \frac{f_{,R}}{2f}\left (f^2 - 3 {r_{\rm h}^{\prime}}^2 \right) \right] dT^2 + \frac{r_{\rm h} f}{\kappa} d \Omega^2
  \label{eq:extrinsic-curvature-kottler}
\end{equation}
where $r_{\rm h}^{\prime\prime} = d^2r_{\rm h}/dT^2$ and all quantities are evaluated at $R = r_{\rm h}(T)$, as found by \cite{Fleury2013} (equations 2.12 and 2.17).

Equating the components of \autoref{eq:extrinsic-curvature-frw} and \autoref{eq:extrinsic-curvature-kottler}, we obtain

\begin{equation}
  \frac{r_{\rm h} f}{\kappa} = a(t)r\sqrt{1-kr^2}.
  \label{eq:kappa-to-fprime}
\end{equation}

Combining \autoref{eq:kappa-to-fprime} with \autoref{eq:kottler-kappa}, we can eliminate $\kappa$. We can also replace $r_{\rm h}^{\prime}$ using the relation obtained in \autoref{eq:r-to-ar}, since

\begin{equation}
  r_{\rm h}^{\prime} = \frac{d(ar)}{dT} = \frac{da}{dt}\frac{dt}{dT}r,
\end{equation}
where $da/dt$ is given by the Friedmann equation \ref{eq:friedmann-equation}. Following through with the algebra, we arrive at the somewhat intuitive result that both regions must have the same cosmological constant $\Lambda$ and that the Kottler hole has to be mass compensating, meaning the enclosed mass $M$ has to equal the mass excised from the FRW background. 

From the junction conditions the rate of expansion of the hole in static coordinates can also be obtained. By combining \autoref{eq:kottler-kappa} and \autoref{eq:kappa-to-fprime}, we get an expression for $r_{\rm h}^{\prime}$

\begin{equation}
  r_{\rm h}^{\prime} = f(r_{\rm h}) \sqrt{1- \frac{f(r_{\rm h})}{1-k r_{\rm h}^2}}.
  \label{eq:hole-expansion-in-kottler-dR-dT}
\end{equation}
This is the rate that the hole is expanding in Kottler coordinates. Inside the hole, this is integrated together with the Kottler geodesics to determine the point where the light ray exits the hole.

The last thing we need from the boundary conditions is to relate the tangent vectors between the two metrics. The continuity of the metric, imposed by the first junction condition, implies that the connection does not diverge across the boundary. To obtain $\dot{R}$ in terms of FRW tangent vectors $\dot{r}$ and $\dot{t}$, we differentiate \autoref{eq:r-to-ar} and substitute $da/dt$ with the Friedmann equation \autoref{eq:friedmann-equation}. Keeping in mind the boundary conditions, we get an expression for $\dot{R}$. The angular coordinates and angular tangent vectors are unchanged when moving from the Kottler to FRW coordinates, and vice versa. With $\dot{R}$ and $\dot{\phi}$, $\dot{T}$ then can be obtained from the null condition. The result is

\begin{subequations}
  \begin{align}
    \dot{T} &= \frac{\dot{t}}{f}\sqrt{1-kr^2} + \frac{a\dot{r}}{f\sqrt{1-kr^2}} \sqrt{\frac{2M}{ar} - kr^2 + \frac{\Lambda}{3}a^2 r^2} \\
    \dot{R} &= \dot{t}\sqrt{\frac{2M}{ar} - kr^2 + \frac{\Lambda}{3}a^2 r^2} + a\dot{r}\\
    \dot{\phi} &= \dot{\phi}\\
    \dot{\theta} &= \dot{\theta}.
  \end{align}
  \label{eq:kottler-to-frw-transform-jacobian}%
\end{subequations}
The quantities above are all evaluated at the boundary of the hole. This result is given for flat space in \cite{Schucker2009} and \cite{Fleury2013}, but here it has been extended to allow for arbitrary spatial curvature. The reverse transformation can be obtained by inverting the Jacobian from above.

\subsection{Conversion to observed quantities}

\begin{figure}
  \centering
  \includegraphics[height=0.4\linewidth]{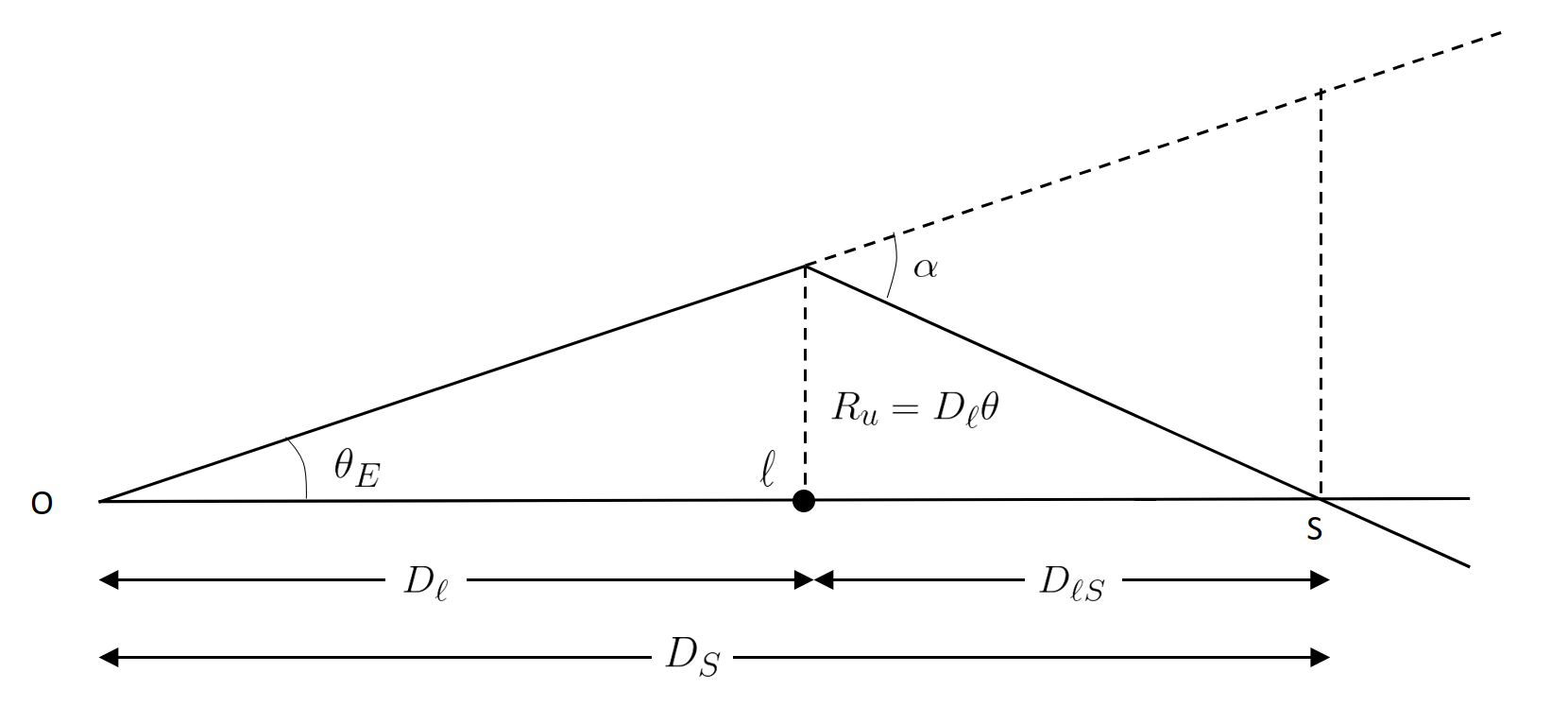}
  \caption{Diagram of gravitational lensing, where the lens, observer, and source are collinear.  Note that in the standard formalism the bend (which in this case occurs only for rays that pass within the vacuole) is assumed to occur in an infinitesimally thin region in the lens plane.}
  \label{fig:lensing}
\end{figure}

Our aim is to compare our results against the standard lensing equations and check whether they agree. 
As indicated,  we consider a simple arrangement where the observer, lens, and source are aligned, as shown in \autoref{fig:lensing}, which is enough to address the light bending question. In the standard lensing analysis, the radial extent of the lens is assumed to be negligible, and the bending is all assumed to happen at a single point as the photon crosses the lens plane. The deflection angle is then defined by

\begin{equation}
  \alpha = \frac{D_{\rm S}\theta_{\rm E}}{D_{\ell {\rm S}}}
  \label{eq:lens-eqn}
\end{equation}
where $D_{\rm S}$ is the angular diameter distance from observer to source, $D_{\ell {\rm S}}$ is the angular diameter distance from lens to source, both of which can be obtained from the numerical integration. We will compare this $\alpha$ with that of the standard Schwarzschild solution, and with the approximate formula of \cite{Kantowski2010}, which includes effects such as the finite volume in which the bending takes place.

The parameter of the unperturbed trajectory $R_{\rm u}$ can also be expressed in terms of the observable quantities $R_{\rm u} = D_{\ell} \theta_{\rm E}$, and $D_{\ell}$ is fixed at the start of the integration. Since we put the lens instead of the observer at the origin, we have to take extra care when converting from raw radial coordinates back into angular diameter distances, especially in curved space. The observer is at  $(r, \phi) = (r_{\ell}, \pi)$, and assuming the source is found through numerical integration to be at $(r, \phi) = (r_{\ell S}, 0)$, we get

\begin{subequations}
  \begin{gather}
    D_{\rm S} = \frac{S_k[S_k^{-1}(r_{\ell {\rm S}}) + S_k^{-1}(r_{\ell})]}{1+z_{\rm S}}\\
    D_{\ell {\rm S}} = \frac{r_{\ell {\rm S}}}{1+z_{\rm S}}.
  \end{gather}
\end{subequations}

This allows us to use \autoref{eq:lens-eqn} to calculate the deflection angle derived from our numerical simulation. We then want to compare this with the expected bending angle according to current standard gravitational formalism. We use the Schwarzschild bending angle, which is expressed in terms of the distance of closest approach $r_0$ in \citep{Keeton2005}, but can be converted into a series expansion in $M/R_{\rm u}$ (Eq. 6 of \cite{Ishak2008}, Eq. 3 of \cite{Butcher2016}) using the relation

\begin{equation}
  \frac{1}{r_0} = \frac{1}{R_{\rm u}} + \frac{M}{R_{\rm u}^2} + \frac{3M^2}{16R_{\rm u}^3}.
  \label{eq:r0-R-relation}
\end{equation}
to give us (to third order in ${M/R_{\rm u}}$):

\begin{equation}
  \alpha_{\text{standard}} = 4 \frac{M}{R_{\rm u}} + \frac{15\pi}{4} \left ( \frac{M}{R_{\rm u}} \right )^2 + \frac{401}{12} \left ( \frac{M}{R_{\rm u}} \right )^3.
  \label{eq:series-expansion-R}
\end{equation}
The numerical integration was done with the \texttt{scipy.integrate.solve\_ivp} function in SciPy\citep{Scipy2020} which uses an explicit Runge-Kutta method of order 5(4) \citep{dormand1980family}. Integration was done separately for each region, starting with the FRW (using \autoref{eq:frw-null-geodesics}), and stops when the boundary of the Kottler hole is reached. The FRW coordinates are converted into Kottler coordinates using \autoref{eq:kottler-to-frw-transform-jacobian}, then the integration proceeds in the Kottler region for the trajectory of the light ray and the comoving boundary (using \autoref{eq:kottler-null-geodesics} and \autoref{eq:hole-expansion-in-kottler-dR-dT}. Integration stops when light reaches the boundary of the hole and we do a similar calculation to transform Kottler coordinates back into FRW coordinates. We then integrate in the remaining FRW region until the light crosses the $x$-axis, to record the radial coordinate and calculate $\alpha$ through \autoref{eq:lens-eqn}. If the light ray misses the hole entirely, then the bending angle is zero. We expect there to be, and do see, deviations from this formula since light bending is only restricted to a finite volume in our simulations, and this will be further discussed in the next section.

\section{Results}

There are a few different factors at play here. In discussing the results of this numerical integration, let us take a step back to look at the specific parts of ray-tracing that have a $\Lambda$-dependence. These are:
\begin{enumerate}
  \item $\Lambda$ dependence in angular diameter distance $D_{\ell}$ as governed by \autoref{eq:angular-diameter-distance-redshit-z}.
  \item The size of the hole. This is governed by \autoref{eq:junction-conditions-mass-volume}. In flat space, increasing $\Omega_{\Lambda}$ implies decreasing $\Omega_{m}$, which corresponds to the matter density of the universe. If we are to keep the mass constant, the hole size would have to increase as we increase $\Omega_{\Lambda}$. 
  \item The rate of expansion of the hole in static Kottler coordinates, given by \autoref{eq:hole-expansion-in-kottler-dR-dT}. The hole has a constant comoving size, but expands in static coordinates. This affects the amount of time light spends inside the Kottler hole.
\end{enumerate}

Of these, (1) is already accounted for in the lensing equation, and (2) and (3) are features of the Swiss-cheese model, which restricts bending to a finite region. These lead to small differences from the standard equations, which assume bending takes place over all space, but they are nothing to do with $\Lambda$.
%cannot be considered to be genuine $\Lambda$ effects. 
By controlling for these effects, we seek to uncover any residual $\Lambda$ dependence.

To deal with (1), we fix the initial angular diameter distance of the lens instead of the redshift. The simulation was done keeping the mass fixed at $M = 10^{13}M_{\odot}$, Einstein angle of $\theta_{\rm E}= 1''$ and the lens at a $D_{\ell} = 1130$ Mpc, which is the angular diameter distance corresponding to a redshift of about $0.5$ at $\Lambda = 0$ in flat space, and then repeated for larger fixed masses up to $10^{17}M_{\odot}$, with corresponding larger angles. For (2), to fix the size of the hole, we use curved space instead to compensate for the change in $\Omega_{\Lambda}$, so that $\Omega_m$ always remains constant. This allows us to keep both the size of the hole and lensing mass $M$ fixed while we change $\Lambda$. 

\begin{figure}
  \centering
  \begin{subfigure}{.49\linewidth}
      \centering
      \includegraphics[height=0.87\linewidth]{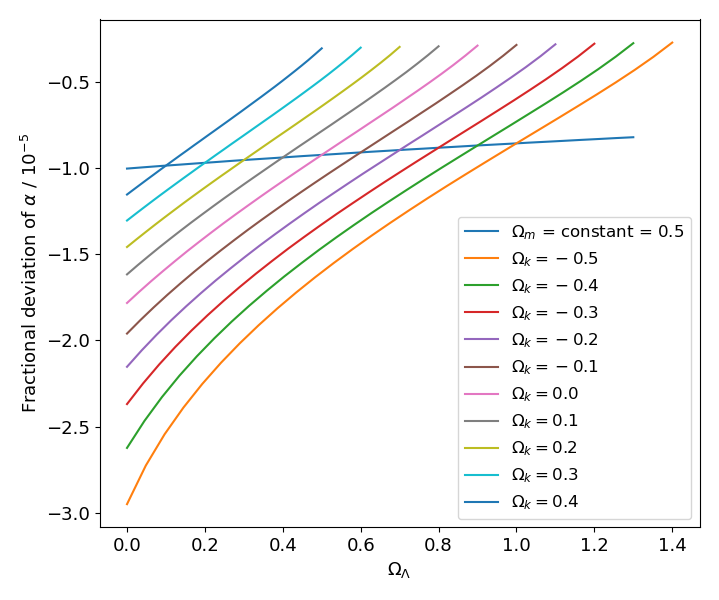}
  \end{subfigure}
  \begin{subfigure}{.49\linewidth}
      \centering
      \includegraphics[height=0.87\linewidth]{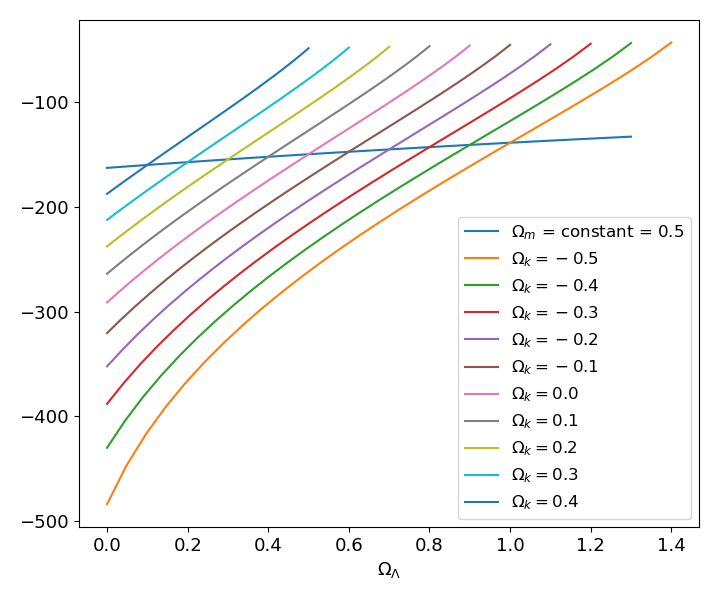}
  \end{subfigure}
  \caption{Plot of the fractional deviations of the deflection angle $\alpha$ from the standard lensing formula \autoref{eq:series-expansion-R}.  On the left we are using $M = 10^{13} M_{\odot}$ and $\theta_{\rm E}=1''$, and on the right $M = 10^{15} M_{\odot}$ and $\theta_{\rm E}=1'$. Lines of $\Omega_k = \text{constant}$ are where $\Omega_m$ was used to compensate for changes in $\Omega_{\Lambda}$, and these exhibit a bigger dependence on $\Lambda$, due to the fact that changing the matter density also changes the hole size. We can see less $\Lambda$-dependent variance on the line where $\Omega_m = \text{constant}$ and $\Omega_k$ was instead used to compensate for changes in $\Omega_{\Lambda}$.}
  \label{fig:deviation-from-frw}
\end{figure}

The deflection angle from the numerical simulation is calculated using \autoref{eq:lens-eqn}. In \autoref{fig:deviation-from-frw}, we plot the the fractional deviation of the numerical result from the expected bending angle (\autoref{eq:series-expansion-R}), $\alpha_{\text{numerical}} / \alpha_{\text{expected}} - 1$. For comparison, we also plot lines of constant curvature, where $\Omega_m$ compensates for the change in $\Omega_{\Lambda}$. We can see that while there appears to be a mass-dependent $\Lambda$ dependence on the order of $10^{-5}-10^{-2}$ for constant curvature, most of this variation goes away when $\Omega_m = \text{constant}$ and the lens mass and size of the hole are fixed. Note that there is an offset even at $\Lambda = 0$, due to the fact that bending happens only in a finite region.

\begin{figure}
  \centering
  \begin{subfigure}{.49\linewidth}
      \centering
      \includegraphics[height=0.87\linewidth]{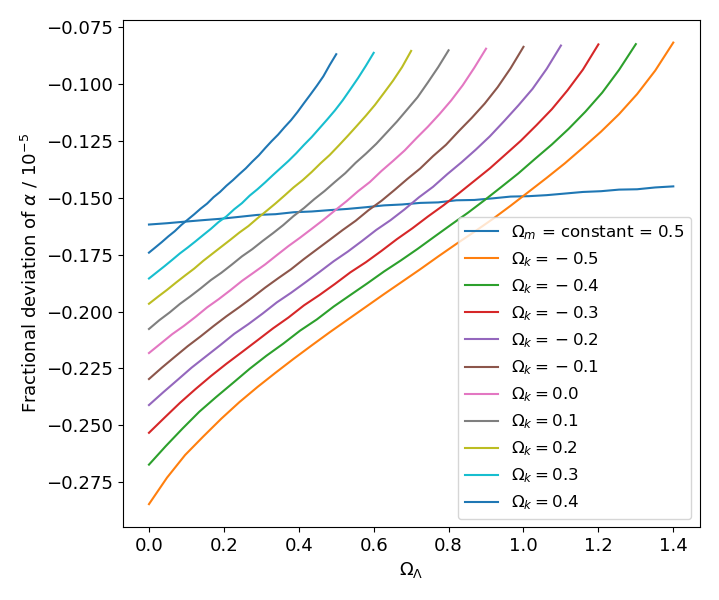}
  \end{subfigure}
  \begin{subfigure}{.49\linewidth}
      \centering
      \includegraphics[height=0.87\linewidth]{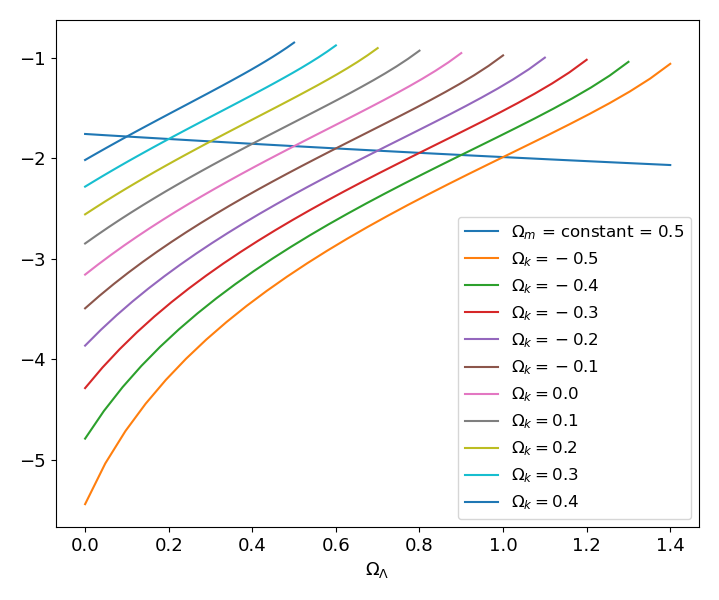}
  \end{subfigure}
  \caption{This plot shows the fractional deviations of the deflection angle $\alpha$ from the predictions in \cite{Kantowski2010}. On the left we are using $M = 10^{13} M_{\odot}$ and $\theta_{\rm E}=1''$, and on the right $M = 10^{15} M_{\odot}$ and $\theta_{\rm E}=1'$. Similar to \autoref{fig:deviation-from-frw}, less deviation is observed for $\Omega_m = \text{constant}$.} 
  \label{fig:deviation-from-kantowski}
\end{figure}

We also compare this result with predictions in \cite{Kantowski2010}, where Kantowski et al. provided equations for how the finite bending range of the Swiss-cheese model affects the bending angle in flat space. We plot the same graph but instead of comparing with expected deflection from conventional lensing, we compare it with the predicted Swiss-cheese bending angle from \cite{Kantowski2010}, so we plot $\alpha_{\text{numerical}} / \alpha_{\text{Kantowski}} - 1$ instead of $\alpha_{\text{numerical}} / \alpha_{\text{expected}} - 1$. As seen in \autoref{fig:deviation-from-kantowski}, for $M=10^{13}M_{\odot}$, our simulations follow the analytic calculations closely, with differences an order of magnitude smaller at around 1 part in $10^6$ that reduce for higher $\Lambda$. 

Since Kantowski et al.'s calculations take into account both the finite size of the hole (effect 2) and the expansion of the hole (effect 3), comparing with their predictions and examining the discrepancies will allow us to take effects 2 and 3 into account. The remaining deviation is much smaller, and is plausibly accounted for by the neglected higher order $\mathcal{O}\left ( 2M/r_0 + \Lambda r_0^2 \right )^{5/2}$ term in \cite{Kantowski2010}, which is of the same order, and also decreases towards higher $\Lambda$.  On the line of varying $\Omega_{\Lambda}$ constant $\Omega_m$, the deviation is almost constant with respect to $\Omega_{\Lambda}$, showing a small upward trend with a fractional change of less than 1 part in ${10^{7}}$.  We suggest that this very small difference may be due to higher order $\Lambda$-dependent changes in the small growth of the hole during the passage of the photon.

\begin{figure}
  \centering
  \begin{subfigure}{.49\linewidth}
      \centering
      \includegraphics[height=0.87\linewidth]{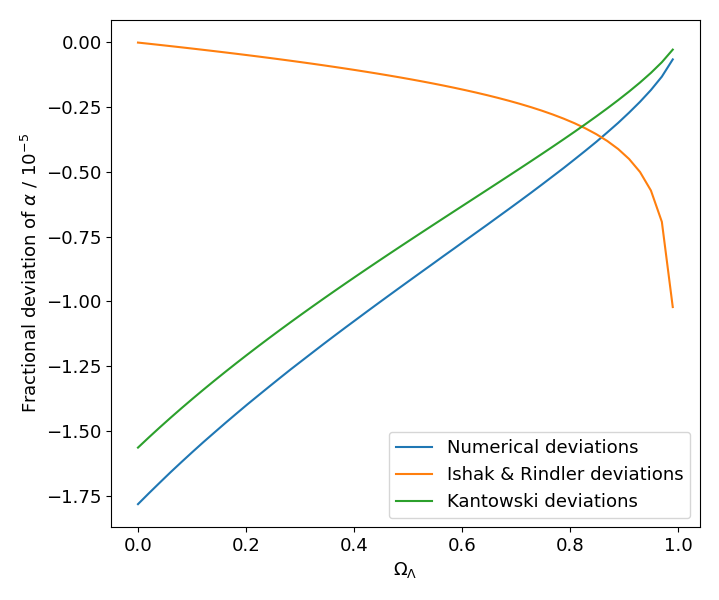}
  \end{subfigure}
  \begin{subfigure}{.49\linewidth}
      \centering
      \includegraphics[height=0.87\linewidth]{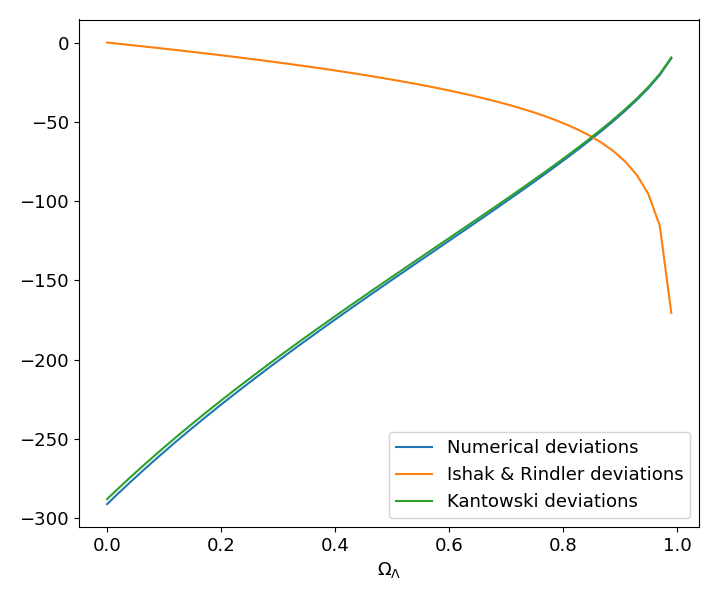}
  \end{subfigure}
  \caption{Plot of the fractional deviations of deflection angle $\alpha$ from standard lensing formula \ref{eq:series-expansion-R} for numerical simulations, Ishak and Rindler predictions \ref{eq:rindler-ishak-2010} from \cite{Ishak2010}, and Kantowski predictions in \cite{Kantowski2010}. On the left we are using $M = 10^{13} M_{\odot}$ and $\theta_{\rm E}=1''$, and on the right $M = 10^{15} M_{\odot}$ and $\theta_{\rm E}=1'$.}
  \label{fig:flat-space-deviations}
\end{figure}

This variance is smaller than Rindler and Ishak's predictions in \cite{Rindler2007}, and in the opposite direction. In \cite{Rindler2007}, it was postulated that there is a correction term of $-\frac{\Lambda R^3}{6M}$. In a later paper \cite{Ishak2010}, the authors found a different term for the Swiss-cheese model, with the deflection angle given by

\begin{equation}
  \alpha = \frac{4M}{R} + \frac{15\pi M^2}{4R^2} + \frac{305M^3}{12R^3} - \frac{\Lambda R r_b}{3}.
  \label{eq:rindler-ishak-2010}
\end{equation}
where $r_b$ is the boundary of the hole. For comparison, we plot our numerical simulations together with these predictions in flat space (\autoref{fig:flat-space-deviations}) with constant $M = 10^{13} M_{\odot}$ and $10^{15} M_{\odot}$, where $\Omega_m$ compensates for the change in $\Omega_{\Lambda}$, since the predictions in other papers mostly assume flat space. As expected, our numerical results follow predictions from \cite{Kantowski2010} most closely, with both getting closer towards the standard lensing formula as $\Omega_{\Lambda}$ increases and matter density decreases, since the bending occurs in a larger region.

In \cite{Schucker2009}, Schucker used a similar method to estimate the bending angle for a lensing cluster at fixed redshift, and found that when a higher $\Lambda$ was used, the bending angle decreased (Table 4 in \cite{Schucker2009}), which appears to be in agreement with Rindler and Ishak's claim that $\Lambda$ attenuates lensing. Guenouche et al. \cite{Guenouche2018} also used a similar formalism and reached the same conclusion. The decrease in bending angle was as much as 10\% when $\Lambda$ increased by 20\%. We were able to reproduce the Schucker's results, but we found that this $\Lambda$ effect can be explained by effects that we are already familiar with and the apparent dependence of bending angle on $\Lambda$ can be explained by effect (1). Given the same redshift, in a flat universe, when $\Lambda$ changes, angular diameter distance is also affected, with the relationship given by \autoref{eq:angular-diameter-distance-redshit-z}. For this case, $D_{\ell}$ increases as $\Lambda$ increases. As the lensing cluster is further away, it makes sense that bending is reduced, which is already quantified in standard lensing formula (\autoref{eq:lens-eqn}). In \autoref{table:schucker-table}, we reproduce a portion of Table 4 in \cite{Schucker2009} with an additional row of the angular diameter distances, and it is easy to see that the deflection angle is inversely related to the angular diameter distances, which is what we expect. 

\begin{table}[ht]
\begin{center}
\begin{tabular}{|c||c|c|c|}
\hline
$\Lambda$&$0.61$&$0.77$&$0.92$ \\ 
\hline\hline
$-\varphi_{\rm S}\ ['']$&10.6 & 10.0&9.0 \\ 
\hline
$M\  [10^{13}M_\odot]$&1.8 & 1.8& 1.7   \\ 
\hline
$D_{\ell} [\text{Mpc}]$&1402&1505&1638 \\
\hline
\end{tabular} 
\end{center}
\caption{Reproducing first 3 columns of table 4 from \cite{Schucker2009}, with an additional row for the angular diameter distance of the lens $D_{\ell}$. $-\varphi_{\rm S}$ is the deflection angle, which appears to decrease as $\Lambda$ increases. However, if we look at the trend of $D_{\ell}$, we can see that the  $-\varphi_{\rm S}$ changes may be accounted for by the changes in $D_{\ell}$ (see Table \ref{table:schucker-table-2}).}
\label{table:schucker-table}
\end{table}

We test this out with a simplified version of Schucker's model, where we kept mass constant, fixed $D_{\ell}$ instead of $z$, and allowed redshift of the source $z_s$ to vary with the integration. We found that when $\Lambda$ increases, the 10\% effect disappears, and only a residual $10^{-7}$ fractional residual effect remain, which is in line with our previous simulations. The result is \autoref{table:schucker-table-2}, where $-\varphi _{\rm S}\ ['']$ is the same for changes in $\Lambda$ up to $10^{-7}$. 

\begin{table}[h]
\begin{center}  
\begin{tabular}{|c||c|c|c|}
\hline
$\Lambda$&$0.61$&$0.77$&$0.92$ \\
\hline
$M\  [10^{13}M_\odot]$&1.8 & 1.8& 1.8   \\ 
\hline
$D_{\ell} [\text{Mpc}]$&1505&1505&1505 \\
\hline\hline
$-\varphi_{\rm S}\ ['']$&9.7369995 & 9.7370019&9.7370039 \\
\hline
\end{tabular} 
\end{center}
\caption{Reproducing results in \cite{Schucker2009} with a fixed $D_{\ell}$ and $M$ instead of redshift.}
\label{table:schucker-table-2}
\end{table}
\section{Conclusion}
\label{sec:conclusion}

Using numerical integration of photon geodesics in a Swiss-cheese model, we have revisited the question of whether the cosmological constant $\Lambda$ affects cosmological gravitational lensing.  In standard lensing analysis, $\Lambda$ affects the bending through its well-known effect on the angular diameter distances, but beyond this we find no effects that could genuinely be attributed directly to $\Lambda$.  There are deviations from the standard lensing equations, which are accounted for to high accuracy by the fact that the bending of light in the Swiss-cheese model is restricted to a finite volume (the Kottler hole), as investigated by \cite{Kantowski2010}. After accounting for this, the residual $\Lambda$ dependence is at the level of 1 part in $10^7$ for $M=10^{13}M_\odot$, and this could plausibly be accounted for by small $\Lambda$-dependent changes in the evolution of the hole during the photon passage across the hole.  In any case, the effects are much smaller than those predicted in \cite{Rindler2007, Ishak2010} and trends in the opposite direction (see Fig.\ref{fig:flat-space-deviations}). We also showed that the apparent large $\Lambda$ effect on deflection angles presented in \cite{Schucker2009} can be ascribed to changes in angular diameter distances and that if you take angular diameter distance differences into account, the effect disappears.  Thus our conclusion is that no modification of the standard cosmological lensing equations are required in the presence of $\Lambda$.
\\

{\bf Acknowledgements:} We would like to thank Pierre Fleury for useful discussions, and the referee for helpful comments.

\bibliographystyle{JHEP}
\bibliography{biblio}

\end{document}